\documentclass[fleqn, usenatbib]{mnras}
\usepackage{bbm}
\usepackage{mathrsfs}
\usepackage{amssymb,amsmath,float}
\usepackage{rotating}
\usepackage{color}
\usepackage{xcolor}
\newcommand\al{\alpha}

\newcommand\gam{\gamma}
\newcommand\de{\delta}
\newcommand\ep{\epsilon}

\renewcommand\th{\theta}

\newcommand\rh{\rho}

\newcommand\ta{\tau}

\newcommand\De{\Delta}

\newcommand\Ph{\Phi}

\newcommand\ie{\emph{i.e.}}
\newcommand\eg{\emph{e.g.}}

\newcommand\beq{\begin{equation}}
\newcommand\eeq{\end{equation}}
\newcommand\bea{\begin{eqnarray}}
\newcommand\eea{\end{eqnarray}}
\newcommand\bal{\begin{align}}
\newcommand\eal{\end{align}}

\newcommand\fr{\frac}

\newcommand\ap{\approx}

\renewcommand\bal{\mbox{\boldmath$\alpha$}}

\title{The binding of cosmological structures by massless topological defects}

\author[R. Lieu]{Richard Lieu
\\
Department of Physics and Astronomy, University of Alabama, Huntsville, AL 35899}

\begin{document}
\label{firstpage}
\pagerange{\pageref{firstpage}--\pageref{lastpage}}
\maketitle

\begin{abstract}
Assuming spherical symmetry and weak field, it is shown that if one solves the Poisson equation or the Einstein field equations sourced by a topological defect, \ie~a singularity of a very specific form, the result is a localised gravitational field capable of driving flat rotation (\ie~Keplerian circular orbits at a constant speed for all radii) of test masses on a thin spherical shell without any underlying mass.  Moreover, a large-scale structure which exploits this solution by assembling concentrically a number of such topological defects can establish a flat stellar or galactic rotation curve, and can also deflect light in the same manner as an equipotential (isothermal) sphere.  Thus the need for dark matter or modified gravity theory is mitigated, at least in part.   
\end{abstract}



\section{Introduction}

The nature of dark matter (DM), defined specifically in this Letter as an unknown component of the cosmic substratum responsible for the extra gravitational field which binds galaxies and clusters of galaxies, has been an enigma for more than a century since the pioneering papers of 
\citet{kap22} 
and 
\citet{oort32}, 
Spitzer observation \citep{morales2022}, and Gaia observation \citep{battaglia2022}
on galactic scales, and 
\citet{zwi33} 
and \citet{chaurasiya2024} 
on the scale of clusters.  Although the laboratory search for DM is an ongoing effort, see \eg~\citet{aal23,ceb23}, alternative theories of gravity were also proposed to enable a considerably smaller amount of matter than is required by Newtonian gravity (and General Relativity in the weak field limit) to produce the same gravitational field strength,  see \citet{noj17} and references therein.

The purpose of this paper is to revisit the logical steps that connect DM existence to the standard theory of gravity.  
It turns out that, in the weak field limit where Newtonian theory suffices, there are two types of impulsive source term $\rh (r)$ for the gravitational Poisson equation, which can lead to an attractive central force as the solution.  
The first has magnitude $\propto 1/r^2$ and is associated with an underlying spherically symmetric mass distribution, while the second has magnitude $\propto 1/r$ and is associated with no underlying mass if space is isotropic.  
The first is evidently Newton's law of universal gravitation, but our interest here is in the second solution because not only has it been ignored, but it can also yield a flat stellar rotation curve in an equipotential environment (apart from a logarithmic increase of $\Phi$ with radius $r$) in the absence of mass.

Of course, the availability of a second solution, even if it is highly suggestive, is not by itself sufficient to discredit the DM hypothesis -- it could be an interesting mathematical exercise at best.  One naturally queries (a) the physical meaning of the singular sources of the type which give rise to the solution; (b) the stability of the solution; (c) the robustness of the model in accounting for a variety of flat rotation curve of galaxies and velocity dispersion in clusters of galaxies; and (d)  observational evidence (beyond the virialized motion of stars in a galaxy and galaxies and hot gas in clusters) of such sources.

We will address (a) in the next section, and (b) in section 3 where we will demonstrate stability by solving the Einstein field equations to show that the metric tensor for the new solution is time-independent.  Next, (c) and (d) are the subjects of sections 6 and 7 where we shall argue that the properties of the massless singular shell sources capable of driving an attractive central force field are specified by several parameters having values to be determined observationally; moreover, 
the increasing frequency of sightings of ring and shell like formation of galaxies in the Universe lends evidence to the type of source being proposed here.  Beyond that, we will also show in section 5 that, in respect of (c), the proposed model can reproduce the bending of light, hence the gravitational lensing
\footnote{Gravitational lensing probes the metric tensor in a manner not shared by the observation of motion of material bodies.} 
by a DM isothermal sphere without necessarily enlisting DM.  


\section{Birkhoff theorem from Poisson equation: gravity on a massless spherical shell in empty space}

We begin with the gravitational Poisson equation as applied to an isotropic environment 
\beq 
\nabla^2 \Phi = \fr{1}{r^2} \fr{d}{dr} \left(r^2 \fr{d\Ph}{dr}\right) = 4\pi G\rho (r). \label{P} 
\eeq 
It is readily seen that for a {\it continuum} mass distribution a radially directed force per unit mass of the form $-\hat r d\Ph/dr = -Gm(r)\hat r/r^2$ is implied by (\ref{P}), where $m(r)$ is the underlying mass, or the integral of $\rho (r)$ over the spherical region of radius $r$.  
Specifically the solution of (\ref{P}) for an inverse-square density source $\rho (r) = A/r^2$ is 
$\Ph (r) = 4\pi G A\log r$,
where the constant $A=m(r)/(4\pi r)$ with $m(r) \propto r$ being the enclosed mass within radius $r$, which leads to the requisite force per unit mass of a flat rotation curve $\mathbf{F} = -\hat r d\Ph/dr = -4\pi GA\hat r/r$.  The amount of source mass $m(r)$ underlying a Keplerian circular orbit of radius $r$ and velocity $v$ is then given by the equation $v^2 = 4\pi GA = Gm(r)/r$, which for a typical observed orbital velocity $v=300$~km~s$^{-1}$ at $r=3$~kpc is a colossal $m = 1.2 \times 10^{44}$~g, almost an order of magnitude larger than the mass of all the observable baryons within such a galactic scale \citep{bland2016,chan2019} 
hence the existence of DM seems to be a justifiable conjecture.  Such a conclusion is inevitable because both the test mass acceleration $v^2/r$ and the mass encumbering beneath the orbital radius are proportional to $A$, hence to each other.

While attempting to look for a way out, consider a discrete density source in the form of a thin spherical shell centred at the origin of total mass $M$, which is a {\it constant}, radius $R>0$ (also a constant) and density 
\beq \rho (r) = \fr{M\delta (R-r)}{4\pi r^2}, \label{den} \eeq 
where $\delta (x)$ is the 1-D Dirac delta function.  The total mass enclosed by a spherical surface of radius $r$ is \beq m(r)=0,~r<R;~{\rm and}~M,~r>R. \label{birk} \eeq   The general solution of (\ref{P}) has two constants of integration.  By choosing them to ensure there is no singularity of $\Phi$ at the origin and $\Phi\to 0$ as $r\to\infty$,  one obtains \beq \Phi(r) = -\fr{GM}{r}~r\geq R,~{\rm and}~-\fr{GM}{R}~r<R. \label{soln} \eeq 
This gives, in essence, the Birkhoff theorem in the weak field limit where General Relativity reduces to (\ref{P}), namely the force field $-d\Phi (r) /dr$ vanishes in the region $r<R$ inside the shell (analogous to Newton's hollow sphere) and is radially attractive outside the shell, being given by the inverse square law with $M$ as the source of gravitational mass.  


At this point, we depart from the conventional approach.  Apart from (\ref{den}), there is at least one other type of shell singularity, or more precisely, an isotropic impulsive source term in the form of a topological defect, capable also of producing a central force field. Consider replacing $\rho (r)$ in (\ref{den}) by a linear combination of $\de (R-r)$ and $\de' (R-r)$, namely \beq \rho(r) = \fr{c^2}{8\pi G} \left[\fr{2\al s}{r^2} \de (R-r) +\fr{2\al s}{r} \de' (R-r) \right], \label{rho} \eeq where $\al>0$ is dimensionless and $s\to 0^+$ is a length, such that \beq s\de (R-r)=1~{\rm at}~r=R. \label{sR} \eeq  In this formulation,
the mass within a sphere of radius $r>R$ is 
\bea 
m(r) &=& \int_0^r 4\pi r^{'2} \rh (r') dr'\notag\\ &=& \fr{\al s c^2}{G} \int_0^r [\de (R-r') + r\de' (R-r')] dr' \notag\\ 
&=& \fr{\al s c^2}{G}\left[ r' \de (R-r') \right]_0^r\notag\\
&=& M\left[ r' \de (R-r') \right]_0^r\notag\\
&=& 0, \label{mass} 
\eea 
where $M=\al s c^2/G$.

Consequently, the singular shell structure contributes nothing to the total mass of the spherical cavity within which it resides.  Thus, the mass of the shell itself must vanish. 

To analyze further the mass distribution, note that part of the source function represents the topological defect of a dipole shell because the derivative of a Dirac delta function at $r=R$ is the limiting case of a half Gaussian for $r<R$ joining with an inverted half Gaussian for $r>R$, with the former corresponding to a thin positive density shell and the latter negative of the same magnitude.  
Since the outer shell has a slightly larger radius, the total mass of both shells ought to be negative, had it not been for the presence of another term in the density function (\ref{rho}), which is proportional to $\de (r-R)$.  
When both terms in (\ref{rho}) are included, the total enclosed mass for any radius $r>R$ vanishes exactly.  

Moreover, as the width of the two Gaussians tends to zero, there is {\it no} finite spherically symmetric region (be it cavity or shell) over which the integral of $\rh (r)$ yields a resolvably negative mass $m(r)$ and its potentially undesirable consequences.  Thus, if one computes the mass of the inner half of the shell where the density is positive by applying (\ref{sR}) and (\ref{mass}) with $r=R$, 
one obtains\footnote{This strictly speaking is the total mass of the central and hollow region $0\leq r<R$ where both the density and mass vanishes, and the inner half of the spherical shell where both are positive.} $m(r=R)=\al Rc^2/G >0$.  But if one lets $r$ exceed $R$ by any finite amount, (\ref{mass}) would give $m(r)=0$ because one has now included the irresolvable outer half of the shell where the density is positive, together with any hollow region which may lie beyond it.   Further discussions on (\ref{rho}) and (\ref{mass}) will take place below.

The potential gradient, on the other hand, is obtained by integrating (\ref{P}) once, with $\rho$ as in (\ref{rho}). with the result is 
an inward acceleration of (or an attractive radial force on) a test particle of unit mass), 
\beq F = -\fr{d\Phi}{dr} = -\fr{\alpha s c^2}{r} \de (R-r) = -\fr{GM}{r} \de (R-r). \label{force} \eeq  
Thus, on the singular shell itself, there is a gravitational inward pull of magnitude $\propto 1/r$, the basic characteristic of a force field capable of driving a color{red} of stellar velocities, even though the shell is massless.  With a sufficient number of such unresolvably closely spaced singular shells, the potential $\Phi$ is effectively a continuous function.
One could then envisage a galaxy or cluster of galaxies, both being populated by discrete clumps of baryons, namely the stars and galaxies, respectively, as well as thermal gas as a virialised ensemble with the force responsible for binding them given by (\ref{force}).

Since the shells themselves are massless, the ensuing total mass of each large-scale structure is simply the mass of the baryons, and the need for enlisting dark matter is mitigated, at least in part.  The shells can be spaced 0.1 AU apart in galaxies and 100 pc apart in clusters (see section 6 for details).  For telescopes with resolving power $\approx 1$~arcsec, stars which are 1 pc or further away from the observer will not reveal the shell spacings, nor will galaxies in a cluster located 1 Gpc or further from the observer reveal the shells in clusters.  

\section{Stability of the second solution}

In order to demonstrate the stability of the model, one must find a time-independent metric tensor solution to the Einstein field equations with the stress-energy tensor being that of pressureless matter of density $\rho (r)$ as given by (\ref{rho}).  
If such a solution exists and has the same physical properties as the previous section in the non-relativistic and weak field limit, \ie~with the metric tensor reducing to the static potential function $\Phi (r)$, it would mean the model is stable.

For steady state and isotropic space, a line element is expressible in the form \beq ds^2 = c^2 g^2 (r) dt^2 - f^2 (r) dr^2 - r^2 (d\th^2 + \sin^2 \th d\phi^2).  \label{metric} \eeq   
The Einstein tensor $G_{\mu\nu} = R_{\mu\nu} - g_{\mu\nu} R/2$ of this metric is 
\beq r^2 G_{00} = 1-\fr{1}{f^2 (r)} + \fr{2}{f^3 (r)} rf'(r);
\label{E00} 
\eeq 
and
\beq r^2 G_{11}= 1-f^2 (r) +\fr{2r g'(r)}{g(r)}; \label{E11} \eeq
and
\beq G_{22}=\fr{r\{f(r)[rg^{''}(r)+g'(r)] -g(r) f' (r) - r f'(r)g'(r)\}}{f^3 (r) g(r)} ; \label{E22} \eeq
and \beq G_{33}=G_{22}\sin^2 \theta. \label{E33} \eeq
We shall consider the limiting case in which the functions $f(r)$ and $g(r)$ are expressible in the form $f(r)=1+\de_1$ and $g(r)=1+\de_2$ where $\de_{1,2}$ are generally $r$-dependent perturbations obeying $|\de_{1,2}| \ll 1$ at least for the range of $r$ of interest.  This depicts the weak field limit where the line element (\ref{metric}) tends to Minkowski as $\de_{1,2} \to 0$.  

Then the simplest way of constructing a flat rotation curve environment, valid to order $\de_{1,2}$ is by writing \beq f(r)=1+\al;~~g(r)= 1+\gamma~{\rm log} (r), \label{fg} \eeq where $\al$ and $\gam \log(r)$ are $\ll 1$, the latter at least within a certain range of $r$ relevant to observations.  When (\ref{fg}) is substituted into (\ref{E00}) through (\ref{E22}) to calculate the Einstein tensor one finds, to order $\al$ and $\gam$,
\beq r^2 G_{00} = 2\al;~~r^2 G_{11} = 2(\gam - \al);~~G_{22}=G_{33}=0. \label{G} \eeq  In the case of pressureless and non-relativistic matter, 
one can ignore fluid pressure $P$ and bulk flow velocities\footnote{$u_i =g_{i\nu} dx^i/d\ta \approx -dx^i/dt = -u^i$ because the lowest order approximation to $f(r)$ and $g(r)$ is $f(r)=g(r)=1$.}  $u_i$ (namely, $P \ll \rho c^2$ where $\rho$ is the mass density of the fluid, and $u_i u^i\ll c^2$); the corresponding stress energy tensor $T_{\mu\nu} = (\rho + P/c^2) u_\mu u_\nu - Pg_{\mu\nu}$ has $T_{00}$ as it's only non-negligible component.  

Thus to lowest order in $\al$ and $\gam$, (\ref{metric}) and (\ref{fg}) form a solution of the Einstein field equations  \beq G_{\mu\nu} = \fr{8\pi G}{c^4} T_{\mu\nu} = \fr{8\pi G}{c^4} \fr{u_{\mu} u_{\rm \nu}}{c^2} \rho \label{E} \eeq  if \beq G_{00}=\fr{8\pi G\rho}{c^2};~{\rm and}~G_{11}=G_{22}=G_{33}=0, \label{rh} \eeq  to this order.  
From (\ref{G}) and (\ref{rho}), one finds that \beq \al=\gam, \label{algam} \eeq in which case \beq G_{00} = \fr{2\al}{r^2} \label{2gamG00} \eeq. 


When the angular momentum of a test particle is finite, the time homogeneity and azimuthal symmetry of (\ref{metric}) allows it to be recast in a form (assuming an orbital plane at $\th = \pi/2$) involving the constants of the motion $\ep=g^2 dt/d\ta$ and $\ell = r^2 d\phi/d\ta$ with $cd\ta=ds$, as \beq \fr{f^2 g^2}{c^2} \left(\fr{dr}{d\ta}\right)^2 +\fr{\ell^2 g^2}{c^2 r^2} +g^2 -1 = \ep^2 -1. \label{L} \eeq In particular, for circular orbits where $r$ is constant and the first term vanishes, differentiating the remaining terms in (\ref{L}) w.r.t. $r$ yields \beq \fr{g'}{g} -\fr{\ell^2}{c^2 r^3} = 0 \implies \al c^2= \fr{\ell^2}{r^2} \implies v=\sqrt{\al} c \label{centrifugal} \eeq in the weak field and low-velocity limit, where $v=\ell/r$ is the (constant) tangential speed of the test particle.  Thus one finds a centrifugal acceleration \beq \fr{v^2}{r} = \fr{\al c^2}{r}. \label{accel} \eeq 
In fact, the constancy of $v$ in (\ref{centrifugal}) accounts for the flat rotation curve of galaxies.      


For a typical speed of $v \approx$~ 300~km$^{-1}$, one requires $\al \approx 10^{-6}$.  This also means the $G_{11}$ tensor component, which is only finite to $O(\al^2)$, lies below $G_{00}$ by a factor $\approx 10^6$, consistent with $G_{11}/G_{00} \approx u_r^2/u_0^2 \approx u_r^2/c^2 \approx 10^{-6}$. This reality check can also be applied to $G_{22}$ and $G_{33}$ leading to the same conclusion.

Is the centrifugal acceleration (\ref{accel})  driven by an amount of underlying mass consistent with Newton's law of universal gravitation, as is to be expected given that $v\ll c$?  To check that, we may use the density $\rho (r)$ of the non-relativistic and pressureless cosmic substratum, as inferred from (\ref{rho}) and the value of $G_{00}$ in (\ref{2gamG00}), to compute the equivalent Newtonian mass\footnote{The term refers to the equivalent total mass of an {\it underlying} (or {\it enclosed}) spherically symmetric matter-energy distribution which produces the same attractive central force (by Newton's theory of universal gravitation) as a shell. As proven earlier, the actual total mass of the shell itself vanishes exactly, even though the attractive force it exerts on a test particle riding the shell is finite and given by (\ref{force}).} $m(r)$ within a circular orbit of radius $r$, namely 
\beq m(r) = \int^r 4\pi r^{'2} \rho(r') dr' = \fr{\al c^2 r}{G} \label{M} \eeq 
after ignoring any factor $f(r)$ in the integral which is O(1) by (\ref{fg}).  
Thus, the Newtonian acceleration is $-Gm(r)/r^2 = -c^2 \al/r$, which is in agreement with (\ref{accel}) and (\ref{centrifugal}). Moreover, comparing to the Poisson equation approach to the same problem, one finds that the two treatments yield the same amount of underlying Newtonian mass equivalent because $\gam$ relates to the constant $A$ in the first paragraph of the last section by the equation $\al = 4\pi GA/c^2$.  The development so far simply spells out the obvious, \ie~ General Relativity reduces to Newtonian Physics in the weak field and non-relativistic limit.  There is no explicit time dependence in the metric, and hence, both the solution and the source are stable.

\section{Flat rotation curve}

What sort of line element in General Relativity would correspond in the weak field limit to the Newtonian (Poisson equation) approach to the $1/r$ attractive force on the massless singular spherical shell described in section 2?   The reader could easily verify that a line element with exactly the same characteristics is given by (\ref{metric}) having \beq f(r)=1+\al s \de (R-r);~~g(r)=1+\fr{\al s\th (r-R)}{R} +\Phi_0, \label{fgshell} \eeq where $\th (r)$ is the Heaviside step function. In (\ref{fgshell}), $g(r)$ differs infinitesimally from unity because $s\to 0$, (\ref{alpha}).  By letting $s$ be the width of the Dirac delta function, \ie~$s \to 0$ such that \beq s \de (R-r) =1~~{\rm at}~r=R. \label{alpha} \eeq  Further, by letting the dimensionless parameter $\al$ satisfy $\al\ll 1$ and restricting oneself to lowest order terms in $\al$ it can readily be shown, using (\ref{E00}) through (\ref{E33}), that
\beq G_{00} = \fr{8\pi G\rho (r)}{c^2}~~G_{11}=G_{22}=G_{33}=0, \label{Gij} \eeq 
where $\rho (r)$ is given by (\ref{rho}).   Thus (\ref{Gij}) is consistent with (\ref{rh}).  In particular, the total mass the shell contributes to a concentric spherical region of radius $r>R$, as given by (\ref{mass}), which is just the integral of $\rho (r)$ from the origin to $r>R$, vanishes.  According to (\ref{accel}), the centrifugal acceleration of a test particle anywhere on the shell itself, namely $r=R$, has magnitude 
\beq  \fr{g' c^2}{g} = \fr{f^2 -1}{2r} c^2 = \fr{\alpha s c^2}{r} \de (R-r)=\begin{cases} \fr{\al c^2}{r},~r=R& \\ 0, ~r\neq R & \\ \end{cases} \label{flatrot} \eeq 
where the last equality is enforced by (\ref{alpha}).  Note that (\ref{flatrot}) is consistent with the equipotential sphere result, namely (\ref{accel}) and (\ref{force}).  The speed of the test particle on the shell is $v=\sqrt{\al} c$, in agreement with (\ref{centrifugal}).
In this way, if there are $N$ concentric singular shells with density satisfying (\ref{rho}) within a galaxy or cluster, \bea &f(r)& = \left[ 1+\al s\sum_{n=1}^{n_L} \de (nR-r)\right];\notag\\ 
&g(r)&=1+\al s \sum_{n=1}^{n_L} \left[\fr{\th (r-nR)}{nR}\right] +\Phi_0, \label{frgr} \eea where in (\ref{frgr}) $\Phi_0$ is a constant but the shells need not be evenly spaced, \ie~$N \leq n_L$ and the $<$ sign is because in the summation sign of (\ref{frgr}) $n$ can change in steps $\de n  > 1$. 


The total mass of the galaxy (cluster) still vanishes
\footnote{The enclosed mass is finite if the spherical surface coincides exactly with a shell, but the probability of coincidence is zero because the shells are infinitely thin.  Rather, such singular shells manifest themselves by the orbital motion of stars and galaxies (and thermal motion of virialized baryons) riding on them, or the deflection of light passing through them. }; 
but its baryonic contents, namely stars (galaxies) and gas, are primarily moving in bound orbits or random walk (\ie~in thermal virialised form) on the shells, giving the impression of a large amount of dark matter hidden within each spherical shell.  The magnitude of the inwardly directed radial acceleration a test particle experiences on-shell is independent of $R$, and according to (\ref{flatrot}) is given by $\alpha c^2/r$ provided  (\ref{alpha}) holds.  The value of $R$ is arbitrary, but for a galaxy is likely to be below the mean spacing between stars.

\section{Gravitational bending of light}

Since evidence of dark matter in galaxies, groups, and clusters come not only from stellar rotation curves and viralised thermal gas but also gravitational lensing \citep{Massey2010,hoe13,Vegetti2023,Zurcher2023}, we examine in this section how the currently proposed configuration of concentric singular spherical shells could possibly bend light in the same manner as an equipotential sphere.  

In general, for the line element of (\ref{metric}), the gravitational inward deflection of a light ray propagating along the $z$-direction and skirting a large-scale structure at impact parameter $a$ as measured in the $y$-direction is, in the weak field limit,  
\beq \theta =\fr{2a}{c^2} \int_{a}^{\infty}~\fr{1}{\sqrt{r^2 -a^2}} \left(\fr{g'}{g} - \fr{f'}{f} \right) dr. \label{def} \eeq  
For a point mass $M$ at the origin where $f=1+GM/r$,~$g=1-GM/r$, (\ref{def}) yields the standard result $\th = 4GM(c^2 a)$.  For the line element of (\ref{fg}), one obtains instead 
\beq \th =\al\pi, \label{iso} \eeq 
which is the constant deflection scenario of an equipotential (isothermal) sphere \footnote{Ignoring the logarithmic increase in $g(r)$ in (\ref{fg}).}.

The question, therefore, is whether the line element of concentric singular shells, (\ref{frgr})
could also lead to a constant inward deflection like (\ref{iso}).  Applying (\ref{frgr}) to (\ref{def}), one finds, after accounting for the oddness of $\de' (x)$, 
\beq\theta = 2\al s\left\{\sum_{n=n_S}^{n_L} \int_a^\infty \left[\fr{a\de (nR-r)}{r\sqrt{r^2 - a^2}} + \fr{a}{\sqrt{r^2 -a^2}}\de'(r-nR) \right] dr \right\}, \label{the} \eeq 
where the smallest shell is indexed $n_S$ and satisfies $n_S R - a \approx R \ll a$; and the largest shell is indexed $n_L$ which satisfies $n_L R \gg a$.  

The first integral corresponds to the $dg/dr$ term in (\ref{def}) and equals $\th_1 = \al\pi s/R$.  The second comes from $-df/dr$ term in (\ref{def}) and may be written after integration by parts as \beq \th_2 = 2\al s
\sum_{n=n_S}^{n_L} \left\{\left[\fr{a\de (r-nR)}{\sqrt{r^2-a^2}}\right]_a^\infty + \left[\fr{anR\th (r-nR)}{(n^2 R^2 -a^2)^{3/2}}\right]_a^\infty \right\} \label{th2} \eeq  where $a\leq nR$.  The first term of (\ref{th2}) vanishes unless the delta function is satisfied at either one of the two limits (it is not satisfied at the upper limit, which lies beyond the large-scale structure), which is improbable for a random sightline because the shells are thin, see (\ref{alpha}).  Thus this term is ignored\footnote{Even though at the lower limit $r=a$ the denominator vanishes, resulting under the $r\neq nR$ scenario in an indeterminate of the form $0/0$, this is resolved by applying L'Hospital rule once to obtain the form $\delta' (r-nR) \sqrt{r^2 -a^2}$ which clearly vanishes at $r=a\neq nR$.}.  The second term is finite only at the upper limit, and equals $2\al s \sum_n anR/(n^2 R^2 -a^2)^{3/2}$.  The summation over $n$ is dominated by the term at which $n^2 R^2 -a^2$ is minimized, namely at the lower $n$ limit (for the given $a$) where $nR \gtrsim a$, or more precisely $nR -a \approx \Delta$ where $\Delta$ is the average distance the light path at the $r=a$ position is from the nearest shell.  Thus the result is \beq \th\approx\th_2 = \fr{\al s}{\Delta} \left(\fr{a}{2\Delta}\right)^{1/2} \label{theta2} \eeq
which is $\gg \th_1$ since $a\gg R$.

In order for the deflection angle $\th \approx \th_2$ to be the constant of the equivalent equipotential sphere (\ref{iso}) one must have, by (\ref{theta2}), 
\beq \Delta = 2^{-1/3} \pi^{-2/3} s^{2/3} a^{1/3}, \label{stoR} \eeq
where it is understood that $s \ll R \lesssim \De \ll a$.  A gradual change in the $s/\Delta$ ratio 
may be interpreted as a mathematical way of enforcing uneven spacing in the sum over shells of (\ref{frgr}), \ie~ both $s$ and $R$ may themselves be constants, but the spacing $\Delta$ between adjacent shells becomes wider towards larger radii because not every integer $n$ corresponds to a shell.
This does not affect the acceleration of stars and galaxies on each shell, as it is determined only by the coefficient $\al$ of the delta function in (\ref{flatrot}).

Before leaving this section about light deflection, it should be noted from the discussion immediately preceding (\ref{theta2}) that although the quantity $nR-a$ in (\ref{th2})  is finite for most line of sights, being given by (\ref{stoR}) on average, there will inevitably be a small fraction of them for which $nR-a \to 0$, resulting in a large $\theta_2$.  This offers a possible discrimination between the current model and standard DM theory; namely, according to the former, the bending of light from background sources by intervening large-scale structures is not totally smooth despite one restricting one's observation to the weak lensing scenario of large impact parameters which take the ray through the structure's outskirts.  That in turn is because if this outermost ray touches one of the shells, it will still be deflected substantially, causing an unusually large tangential shear of the image of the background source, even possibly magnification (\ie~strong lensing).  DM models do not predict such a possible behaviour.  Detailed mapping of background sources behing the outer parts of a cluster or galaxy, in search of large convergence fluctuations there, could offer a scrutizing test.

\section{Role of the baryons}

How may one expect actual rotation curves, as predicted by the shell model, to look like when baryons are taken into account?  On galactic scales, the baryon to DM mass ratio is below the cosmological average of $\ap$~20 \%, \cite{fuk98}, and most of the baryons are in stars if one considers only the galactic disk by ignoring the halo, \cite{cau20}.  Since the model adopted here is spherically symmetric, with applications also to clusters of galaxies\footnote{In clusters the baryon fraction is quite close to the cosmological value, \cite{fuk98}.}, the halo gas has to be taken into account.  For a galaxy like the Milky Way, the inclusion of halo gas would double the stellar mass, and leads to a total baryonic mass of $\approx$~10 \% of the DM mass, \cite{cau20}.  

In this section, we estimate the effect of baryons on galaxy rotation curves in the context of the shell model, by assuming the baryons as a gas in hydrostatic equilibrium in the potential field of the massless shells, with both the gas distribution and the shells being spherically symmetric.   Under this scenario, two equations govern the two unknowns $\rh_b (r)$ and $T(r)$, respectively, the density and temperature profile of the baryons, \cite{bul10}.  They are \beq \fr{1}{\mu m_p n_e(r)} \fr{dP}{dr} =-\fr{d\Phi}{dr};~n_e (r)=n_{eS} \left(\fr{T}{T_S}\right)^p, \label{hydro} \eeq where $n_e$ is the number density of electrons in the halo gas, which is treated here as a plasma of mean molecular mass $\mu$, $P=P(r) = n_e (r) kT(r)$, $m_p$ is the proton mass, $p>0$ is the polytropic index of the plasma, and the subscripts $S$ and $L$ refers respectively to the smallest and largest radii within which the hierarchical shell structure exists (\ie~in the notation of (\ref{frgr}) we have $r_S=n_S R$ and $r_L= n_L R$).  

Now in (\ref{frgr}) one may in the limit of many shells, approximate the summation by an integral, namely \beq g(r) = 1+\Phi (r), \label{gofr} \eeq with \bea \Phi(r) &=& 
2^{1/3} \pi^{2/3} s^{1/3} \al c^2 \int_{r_S}^r \xi^{-4/3} d\xi +\Phi_0 \notag\\
&=& -3\times 2^{1/3} \pi^{2/3} \al c^2 s^{1/3} r^{-1/3}.\label{potential} \eea In (\ref{potential})  $\xi =nR$, and use was made of (\ref{stoR}).  Specifically, due to the change in the shell index $n$ becoming $\de n > 1$ towards larger radii, fewer shells are summed compared to the regular spacing scenario of $\de n =1$.  This means, starting with the continuum representation of the regular spacing case, \beq \Phi (r) = \al s c^2 \sum_{n=n_S,1}^{n_L} \fr{\th (r-nR)}{nR} + \Phi_0, \label{regular} \eeq one must now replace $d\xi$ by $Rd\xi/\Delta < d\xi$, which is what we did (\ref{potential}).   Moreover, in arriving at the last step of (\ref{potential}) we fixed the constant $\Phi_0$ such that $\Phi (r) \to 0$ as $r\to\infty$.  Note that the shells must share a common acceleration parameter $\al$ to yield a flat rotation curve for stars in a galaxy.
The continuum approximation in (\ref{potential}) means the baryonic gas is able to occupy the space in between shells by collision and subsequent virialization, while stars are bound to the shells and follow circular orbits,

The three equations (\ref{frgr}), (\ref{gofr}),  and (\ref{hydro})  may then be solved to yield
\beq T(r)=-\fr{1}{1+p}\fr{\mu m_p}{k} \Phi(r) =\fr{3\times 2^{1/3} \pi^{2/3} \al c^2 s^{1/3}}{1+p} \fr{\mu m_p}{k} \fr{1}{r^{1/3}}; \label{tempr} \eeq
and
\beq M_b (r)= \fr{12\pi\mu n_{\rm eS} m_p r_S^{p/3}}{9-p} \left[ r_L^{(9-p)/3} - r_S^{(9-p)/3}\right],~p\neq 9.
\label{Mb} \eeq  Such an amount of baryons would provide an extra acceleration $-GM_b (r)\hat{r}/r^2$ which is to be added to the shell acceleration of (\ref{flatrot}) to obtain the total circular rotation speed of a star, due to the shell it is on and the baryons underlying it.  Results for various polytropic indices $p$ are shown in Figure 1, where it can be seen that a variety of observed rotating curves (see Figure 1 of \cite{zha19}) are reproduced. 

Concerning the actual values of the shell thickness $s$ and spacing $R$, one could conjecture for galaxies that $s=4 \times 10^9$~cm, and $R=3 \times 10^{12}$~cm.  This would satisfy (\ref{stoR}) with $n\approx 10^9$ out to a radius $a\approx nR \approx 1$~kpc.  
The actual shell separation at 1 kpc radius is, by (\ref{stoR}), $\De\approx 1.34 \times 10^{13}$~cm, which is 4 times larger than $R$.  
Thus, at this radius, there is one shell every $\De n \approx 4$.    For $p=3/2$ and $\mu=1$ the temperature of the gas from (\ref{tempr}) is $T_S \approx 4\times 10^3$~K, and the electron density is obtainable from the second of (\ref{hydro}) as $\approx n_{\rm eS} \ap 0.127$~cm$^{-3}$, both evaluated at $r_S = 0.5$~kpc and with the latter being the value which yields the cosmological baryon to DM mass ratio of 18 \% at the radius of 50 kpc.  Such values of $T_S$ and $n_{\rm eS}$ are consistent with our understanding of the warm diffuse HII component of the interstellar gas, \cite{cox05}.  In the case of clusters of galaxies, a pair of galactic rather than solar system scale lengths, such as $s=3\times 10^{19}$~cm (or 10 pc) and $R=2\times 10^{20}$~cm, would by (\ref{stoR}) imply $\ap 1,500$ shells out to the radius $a\approx nR \approx 100$~kpc, at which point the actual separation between consecutive shells is $\De= 2.4\times 10^{20}~{\rm cm} \approx R$ by (\ref{stoR}).   Moreover, for $p=3/2$ a temperature $T_S \approx 1.75\times 10^7$~K for the baryons at $r_S= 100$~kpc radius, by (\ref{hydro}) with\footnote{Clusters have about 10 times higher $\al$ than galaxies because their member galaxies have a velocity dispersion $\approx 1,000$~km~s$^{-1}$ which is $\approx 3$ times higher than galaxies, see \cite{kra12}.  Hence $\al =10^{-5}$ by (\ref{centrifugal}).}  $\al =10^{-5}$.   Moreover, the baryonic mass between $r_S=100$~kpc and $r_L =1$~Mpc is, by (\ref{Mb}) with $n_{\rm eS} =10^{-3}$~cm$^{-3}$, $M_b =3.58\times 10^{13}~M_\odot$ for $p=3/2$.  This is close to the cosmological baryon mass budget of $\ap 18$\% of the DM mass where, by (\ref{M}), $M_{\rm DM} = \al c^2 r_L/G$ with $\al\approx 10^{-5}$ corresponding, via (\ref{centrifugal}), to a velocity dispersion of $v \approx 10^3$~km~s$^{-1}$ which is typical of clusters, as are $n_{\rm eS}$ and $T_S$ (see \eg~\cite{kra12}).

Lastly, as already pointed out in section 2, neither the shell thickness $s$  nor the spacing $\De$ between adjacent shells is resolvable by current telescopes.  Moreover, the value of $s$ is sufficiently small that stars in a galaxy only experience an infinitely thin shell, same for galaxies in a cluster.  
In both cases, they cannot respond to the fine details of the double layer (of positive and negative mass) of a shell.  

\begin{figure}
    \centering
    \includegraphics[width=1\linewidth]{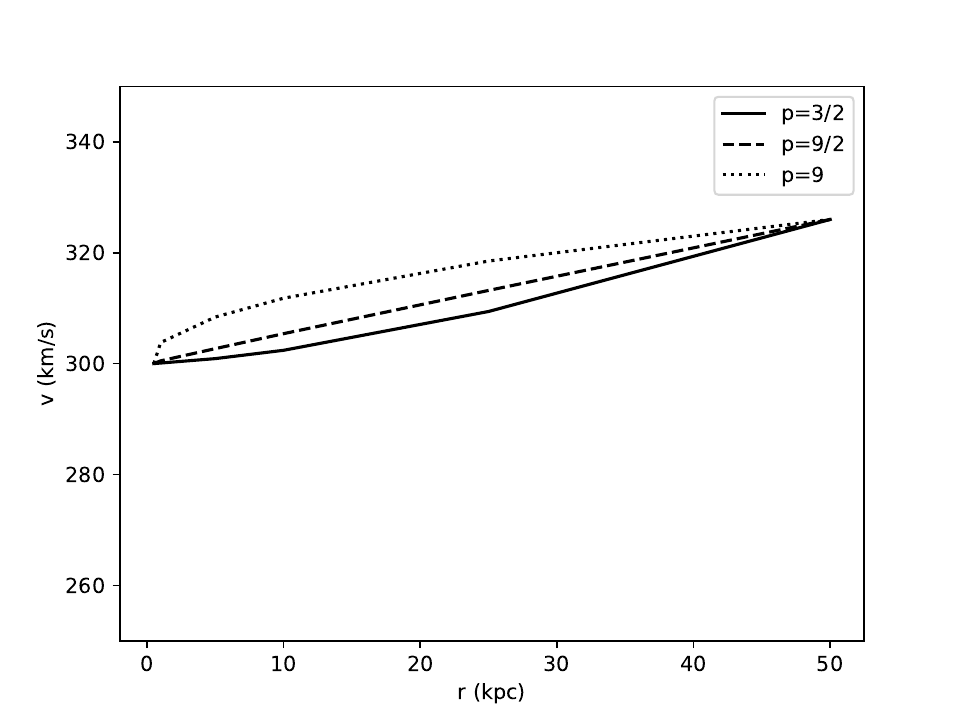}
    \caption{Circular velocity of stars in a galaxy, taking into account the extra centrifugal acceleration provided by the gravitational pull of the underlying baryons, as well as the singular shell.  Here $p$ denotes the polytropic gas index of the baryons, and the central electron density $n_{\rm eS}$ is chosen to enable the baryons within 50 kpc radius to equal 18 \% of the dark matter thereby mass, consistent with the cosmological value of this ratio.  Note that this way of fixing $n_{\rm eS}$ overestimates the baryons in clusters (see text for elaboration), hence the extra acceleration due to the baryons.  Nevertheless, the rotation curve is still sufficiently flat, indicating that the baryons alone do not play much role in providing the necessary centrifugal acceleration. }
    \label{fig:Verlocity_R}
\end{figure}

\section{Evidence for the existence of the proposed topological defects}

Although topological effects, or singular distribution of mass-energy in space, have by now been proposed in many and varied forms since they were first envisaged by \citet{kib76}, so long as they are stable, not all of them are understood in terms of their origin.  An example is a thin sheet of negative mass wrapping around a wormhole to maintain cylindrical symmetry -- the system is shown to be stable -- irrespective of origin, \citet{eir16}.  

As far as the monopole-dipole shell combination in this work is concerned, although its origin is uncertain (despite being deemed stable in section 3), there are scalar field solutions which can give rise to it (see \cite{yut02}), and there is a body of evidence indicative of its existence,
namely the discovery of a giant arc of proper size $\approx 1$~Gpc at $z\approx 0.8$ which adds to an already existing set of cosmologically sized distribution of galaxies in the form of walls and rings, see \citet{lop22} and Table 1 therein.  
These galaxy arcs, rings, and walls appear to be particularly singling out our proposed model because, on Gpc scales, there is simply no way of attributing to DM overdensity as the source of any centrifugal force which could bind together structures of such enormity, \ie~ the Universe is too homogeneous by then.  Thus, the specific topological defect proposed in this paper is reasonably corroborated by the unusually large number of observations of organized shell-like or (as a projection effect) ring-like manifestations of stars and galaxies on kpc scales and beyond.  Such defects can in principle provide the necessary centrifugal force without enlisting `missing' mass.      

\section{Conclusion}

In addition to the well-known Green function solution of the gravitational Poisson equation, which gives rise to no net force inside a thin spherically symmetric shell of mass and an attractive inverse-square force outside, the existence of another singular shell solution which sources no mass but drives an attractive $1/r$ force on the shell itself is demonstrated in this letter.  When a galaxy (cluster) comprises many concentric singular shells of this type, stars (galaxies) orbit, and diffuse baryonic gas can be virialised, on each of the shells.  The result is a flat rotation curve with large Keplerian velocities in the case of orbital motion and high temperature thermal motion in the case of visualized gases; both phenomena are symptomatic of dark matter even though such matter may not exist or exist in an amount smaller than is required for binding a galaxy or cluster.  

The observation of giant arcs and rings (\eg~\cite{lop22}) could lend further support to the proposed alternative to the dark matter model.  It should be pointed out that the thin shells advocated here do not have to cover an entire spherical surface to be effective, \ie~the result of section 2 is readily shown to apply also to a topological defect which occupies part of a spherical surface, although it cannot account for elliptical orbits of a test mass.  In this paper, we also discussed how
gaseous baryons can be in hydrostatic equilibrium with the potential established by the shells, leading to temperature and density in agreement with observations.  Moreover, their `feedback' effect on the acceleration provided by each shell was estimated and was shown to be slight, namely it is only a small perturbation of the shell acceleration of stars and galaxies.  
Hence, the flat rotation curve of stars is not distorted by much.

Apart from rotational curves and virialisation, this paper also addresses the other major piece of evidence for dark matter in large-scale structures -- gravitational lensing.  As it turns out, the proposed concentric singular shells can deflect the light path to a distant point source by a small and constant angle independently of the impact parameter in the weak lensing limit, provided the spacing $\De$ between consecutive shells widens with radius $r$ as $\De \propto r^{1/3}$.  Thus, again, the need to enlist dark matter is not inevitable.

Nevertheless, this paper does not attempt to tackle the problem of structure {\it formation}; nor does it argue against the role of dark matter in accounting for the ratio of the first two acoustic peaks of the cosmic microwave background \citep{roo12}, and the presence of trace Deuterium
\footnote{Safe to say that, like baryons and dark matter, an early Universe comprising only baryons and shell singularities of the type presented here would probably not provide enough collisions to annihilate all deuterium.} 
from Big Bang Nucleosynthesis \citep{kaw15,sib20}.  Rather, we focussed our attention on the evidence as provided by galaxies and clusters of galaxies.

\section{Data availability statement}

No new data were generated or analysed in support of this research.




\label{lastpage}
\end{document}